\documentclass{article}

\usepackage[accepted]{icml2024}

\usepackage{microtype}
\usepackage{graphicx}
\usepackage{subfigure}
\usepackage{booktabs} 

\usepackage[]{xcolor}
\usepackage{url,hyperref}
\definecolor{light_grey}{rgb}{0.3,.3,.3}
\definecolor{deep_blue}{rgb}{0,.2,.5}
\definecolor{dark_blue}{rgb}{0,.15,.5}
\definecolor{myblue}{rgb}{.01,0.21,0.71}
\definecolor{dark_green}{rgb}{0.2,0.5,.3}
\definecolor{firebrick}{rgb}{0.75,0.125,0.125}
\definecolor{indigo}{rgb}{0.3,0,0.5}
\definecolor{dark_brown}{RGB}{85, 39, 0}
\definecolor{deep_brown}{RGB}{128, 70, 21}

\definecolor{lightgreen}{RGB}{138,255,138}
\definecolor{lightblue}{RGB}{138,138,255}
\definecolor{lightorange}{RGB}{255,212,138}

\colorlet{deep_color}{deep_blue}
\colorlet{dark_color}{dark_blue}
\colorlet{mycolor}{myblue}

\hypersetup{pdftex,  
  breaklinks=true,  
  colorlinks=true,
  urlcolor=deep_brown,
  linkcolor=deep_blue,
  citecolor=dark_green,
  }

\usepackage{snaptodo}

\snaptodoset{block rise=2em}
\snaptodoset{margin block/.style={font=\tiny}}

\usepackage{hyperref}

\usepackage[capitalize,noabbrev]{cleveref}

\usepackage[textsize=tiny]{todonotes}

\usepackage[normalem]{ulem}

\icmltitlerunning{Hype, Sustainability, and the Price of the Bigger-is-Better Paradigm in AI}

\begin{document}

\twocolumn[
\icmltitle{Hype, Sustainability, and the Price of the Bigger-is-Better Paradigm in AI}

\icmlsetsymbol{equal}{*}

\begin{icmlauthorlist}
\icmlauthor{Gaël Varoquaux}{equal,inria}
\icmlauthor{Alexandra Sasha Luccioni}{equal,hf}
\icmlauthor{Meredith Whittaker}{equal,signal,uwa}
\end{icmlauthorlist}

\icmlcorrespondingauthor{Gaël Varoquaux}{gael.varoquaux@inria.fr}

\icmlaffiliation{inria}{Inria, Saclay, France}
\icmlaffiliation{hf}{Hugging Face, Montreal, Canada}
\icmlaffiliation{signal}{Signal, NYC, United States}
\icmlaffiliation{uwa}{University of Western Australia, Perth, Australia}

\icmlkeywords{Machine Learning, ICML}

\vskip 0.3in
]

\printAffiliationsAndNotice{\icmlEqualContribution} 

\begin{abstract}
With the growing attention and investment in recent AI approaches such as large language models, the narrative that the larger the AI system the more valuable, powerful and interesting it is is increasingly seen as common sense. But what is this  assumption based on, and how are we measuring value, power, and performance? And what are the collateral consequences of this race to ever-increasing scale? Here, we scrutinize the current scaling trends and trade-offs across multiple axes and refute two common assumptions underlying the ‘bigger-is-better’ AI paradigm: 1) that performance improvements are driven by increased scale, and 2) that all interesting problems addressed by AI require large-scale models. Rather, we argue that this approach is not only fragile scientifically, but comes with undesirable consequences. First, it is not sustainable, as, despite efficiency improvements, its compute demands increase faster than model performance, leading to unreasonable economic requirements and a disproportionate environmental footprint. Second, it implies focusing on certain problems at the expense of others, leaving aside important applications, e.g. health, education, or the climate. Finally, it exacerbates a concentration of power, which centralizes decision-making in the hands of a few actors while threatening to disempower others in the context of shaping both AI research and its applications throughout society.
\end{abstract}

\section{Introduction: AI is overly focusing scale}
\label{sec:intro}


Our field is increasingly dominated by research programs driven by a race
for scale: bigger models, bigger datasets, more compute. 
Over the past decade, machine learning (ML) has been used to build systems used by millions, carrying out tasks such as
automatic translation, newsfeed raking, and targeted advertising. 
Pursuing scale has helped
improve the performance of ML models on benchmarks for many tasks, with the paradigmatic
example being the ability of large language models (LLMs) to encode
``general knowledge'' when pretrained on large amounts of text data. The perceived success of these approaches has further entrenched the assumption that bigger-is-better in AI.
Here, we examine the underpinnings of this assumption about scale and its collateral consequences.
We argue not that scale isn't useful, in some cases, but that there is too much focus on scale and not enough value on other research.

\begin{figure*}[t!]
    \begin{minipage}{.3\linewidth}
    \caption{\textbf{An explosion in model size} --
    Left: The increase in model size means it is more and more 
    expensive to run them in terms of RAM.
    Right: resources we need are increasing faster than available compute.
    Data from \citet{epochMachineLearningData2022}, specific details in \autoref{app:historical_plots}.
    \label{fig:trends_scale}}
    \end{minipage}%
    \hfill%
    \begin{minipage}{.33\linewidth}
	\includegraphics[width=\linewidth]{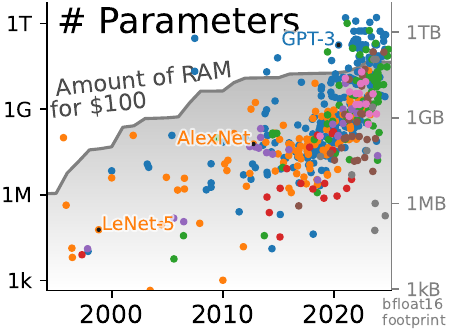} 
    \end{minipage}%
    \hfill%
    \begin{minipage}{.33\linewidth}
	\includegraphics[width=\linewidth]{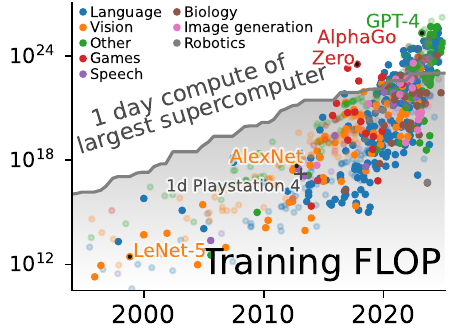}%
    \end{minipage}%
\end{figure*}

\paragraph{From narrative to norm}

The famous AlexNet paper \cite{krizhevsky2012imagenet} has been key
in shaping the current era of AI, including the assumption that increased scale is the key to improved performance. While building on and acknowledging
decades of work, AlexNet created the recipe for the current bigger-is-better paradigm in AI, combining GPUs, big data (at least for the time), and
large-scale neural network-based approaches. The paper also demonstrated that GPUs\footnote{GPUs, or Graphical Processing Units, were initially developed for video games, enabling better graphics rendering given their ability to carry out more processes in parallel compared to CPUs, or Central Processing Units, which had been the dominant hardware architecture.}  scale
compute much better than CPUs, enabling graduate students to rig a hardware setup that produced a model which outcompeted those trained on tens of thousands of CPUs. The AlexNet authors viewed scale as a key
source of their model's benchmark beating performance: ``\emph{all of our experiments suggest that our results
can be improved simply by waiting for faster GPUs and bigger datasets to
become available}''~\citep[p.2]{krizhevsky2012imagenet}.
This point was further reinforced by Sutton in his ``bitter lesson'' article, which states that approaches based on more computation win in the long
term, as the cost of of compute decreases with time~\cite{sutton2019bitter}. 
With this assumption comes both an explosion in investment in large-scale AI models and a concomitant spike in the size of the notable (highly cited) models. Generative AI, whether for images or text, has taken this assumption to a
new level, both within the AI research discipline and as a component of the popular `bigger-is-better' narrative surrounding AI --  this, of course, has impacted training and inference compute requirements, which have ballooned in recent years 
(see \autoref{fig:trends_scale}).

The bigger-is-better norm is also self-reinforcing, shaping the AI research field by informing
what kinds of research is incentivized, which questions are asked (or remain unasked), as well as the relationship between industrial and academic actors. This is important because science, in AI as in other disciplines, does not happen in a vacuum -- it is built upon relationships and a shared pool of knowledge, in which prior work is studied, incorporated, and extended. Currently, a handful of benchmarks define how ``SOTA'' (state-of-the-art) is measured and understood, and in pursuit of the goal of improving performance on these benchmarks, scale has become the preferred tool for achieving progress and establishing new records. Reviewers ask for experiments at a large
scale, both in the context of new models and in the context of measuring performance against existing models \citep[e.g.][]{reviewer2024iclr}; scientific best practices call for running
experiments many times \emph{e.g.} for
hyperparameter selection \cite{bouthillier2021accounting}. These incentives and norms contribute to pushing computing budgets beyond
what is accessible to most university labs -- which in turn makes many labs increasingly dependent on close ties with industry in order to secure such access~\cite{Abdalla_2021,whittaker2021steep,Abdalla_2023}. Taken together, we see the ``bigger-is-better" norm in AI creating conditions in which it is increasingly difficult for anyone outside of large industrial labs to develop, test, and deploy SOTA AI systems. 

The bigger-is-better assumption is also prevalent beyond the AI research community. It is shaping how AI is used and the understandings and expectations about its capabilities. Popular news reporting assumes larger
amounts of compute result in and equate to better results and commercial success~\cite{wodecki2023want,lawmeta2024}. Regulatory determinations and thresholds
assume larger means more powerful and more dangerous, \emph{e.g.} the US Executive
Order on AI~\cite{biden2023executive} and EU AI Act~\cite{parliament2023eu} -- this broader assumption shapes markets and policy-making. 

\paragraph{Paper outline}
We start by discussing why this fixation on scale is
misguided -- we first examine this assumption in terms of how, and whether, scale leads to improvement in goal setting, finding that scale is not well correlated with better performance in certain contexts, and in fact benefits of scale tend to saturate.
(Section \ref{sec:factors}). Then we look at harmful consequences that result from the growth of large-scale AI. Firstly, large scale development is unsustainable (Section \ref{sec:unsustainable}), and the drive for more and more data encourages unethical and unauditable data practices (Section \ref{sec:data}). Further, due to the expense and scarcity of hardware and talent required to produce large-scale AI, bigger-is-better strategies increasingly concentrate power over AI in the hands of a narrow set players (Section \ref{sec:narrowing}). We  conclude by outlining how the research community can reclaim the scientific discourse in the AI field, and move away from a singular focus on scale.

\section{What problems does scale solve?}
\label{sec:factors}

\subsection{Scale is one of many factors that matter}

\begin{figure*}[t!]
    \begin{minipage}{.33\linewidth}
    \caption{\textbf{Performance as a function of scale saturates} across
    various tasks. Plots of performance as a function of scale
    (time or memory footprint) on benchmark data from {\sffamily a}) a medical image segmentation challenge \cite{flare2023challenge}, {\sffamily b})
    computer-vision object detection \citep[COCO]{lin2014microsoft}
    and {\sffamily c}) scene parsing
    \citep[ADE20K]{zhou2017scene},
    {\sffamily d})
    tabular learning \cite{grinsztajn2022tree}, 
    {\sffamily e}) text embedding \citep{muennighoff2022mteb}, and
   {\sffamily f}) text understanding \citep{open-llm-leaderboard}.
    \label{fig:benchmarks}%
    Details in \autoref{app:benchmark_plots}.
    }
    \end{minipage}%
    \hfill%
    \begin{minipage}{.64\linewidth}
	\includegraphics[width=\linewidth]{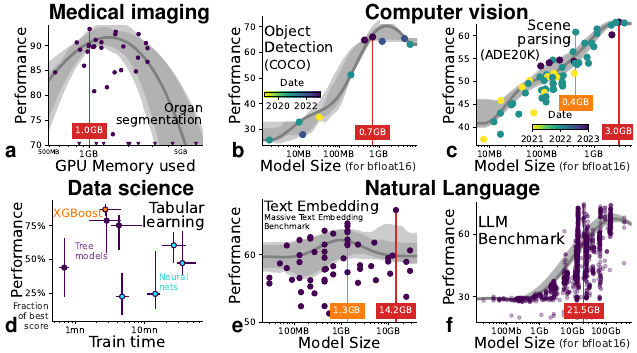}
    \end{minipage}%
\end{figure*}

\paragraph{A staggering increase in scale, and cost}

The scale of notable (highly-cited) models has massively increased over
the last decade, driven by a super-exponential growth in terms of number of
parameters, and amount of compute used
(\autoref{fig:trends_scale}). This growth is much more rapid than the
increased capacity of hardware to execute necessary processing for model training and tuning. Indeed, while the size of large models, as measured by number of parameters, is currently doubling 
every 5 months (\autoref{sec:doubling_time}), the cost of 1 GB of memory has been nearly constant
for a decade. This means that the resources required participate in cutting-edge AI research have increased significantly. The compute used to train an AI model went from a single day on a gaming console (in 2013) to surpassing the largest supercomputers on Earth (in 2020). A common response to concerns about the exponential growth of compute requirements is a reference to Moore's law\footnote{Moore's law is an observation that states that the number of transistors on a microchip doubles roughly every two years.} -- i.e. that computational power will also increase, and this will help ensure that compute remains accessible \citep[see][]{sutton2019bitter}. However, this is not true in practice, since compute requirements for SOTA models are surpassing improvements in computational power. For instance, \citet{thompson2022computational} carried out a meta-analysis of 1,527 research papers from different ML sub-domains and extrapolated how much computational power was needed to improve upon common benchmarks like ImageNet and MS COCO. They estimated that an additional 567 times more compute would be needed to achieve a 5\% error rate on ImageNet, stating that \emph{``fundamental rearchitecting is
needed to lower the computational intensity so that the scaling of these problems becomes less onerous"}. 

Empirically, Sutton's ``bitter lesson'' \cite{sutton2019bitter}
 appears partly incorrect: it is not that, for AI,
``\emph{general methods that leverage computation are ultimately the most effective,}  [because of] \emph{Moore's law,} [...]
\emph{continued exponentially falling cost per unit of computation}'', but that increasing resources are spent on AI.
This increase in resources is visible in computational costs but is also true of other costs. For instance, building larger AI models require more human labor (\autoref{app:labor}).

\paragraph{Diminishing returns of scale}

The problem with the bigger-is-better approach is not simply that it is inaccessible. It also does not consistently produce model improvements, and, after a given point, even shows diminishing returns. On many tasks, benchmark performance as a function of
scale tends to saturate after a certain point (see \autoref{fig:benchmarks}). In addition,
there is as much variability in model performance between models within a similar size class as there is between models of different sizes~\cite{open-llm-leaderboard}.
Indeed, many factors beyond scale are important to produce performant AI models. Choosing the right model architecture for the data at hand is crucial. Transformer-based models, widely perceived to be SOTA on most ML benchmarks, are not always the most fitting solution. 
For instance, when working with tabular data of the kind commonly produced in enterprise environments, tree-based models produce better predictions, \emph{and} are much
faster (and thus less expensive and more accessible) compared to neural network approaches (\autoref{fig:benchmarks}d). This is notably due to the fact that their inductive bias is adapted to the specificity of
columnar data \citep[]{grinsztajn2022tree}.

Even for neural networks, or attention-based models, progress has been driven by much more than scale.
Training or
fine-tuning strategies play a very important role
\cite{davidson2023ai}. For instance, in text embeddings 
across models of the same size,
and often even for the same type of pre-trained model,
fine-tuning to produce useful
similarities can markedly improve the resulting embeddings on domain-specific tasks \citep[][and
\autoref{fig:benchmarks}e]{muennighoff2022mteb}. We see this exemplified in the E5 model, which leverages both contrastive learning and curated data to achieve remarkable results \cite{wang2022text}. 
Considering text-based models, encoder-based models often lead to representations that facilitate learning or language processing much better than decoder-based models, and do so in a much less compute intensive way \cite{grinsztajn2023vectorizing,warner2024smarter}.

\subsection{In many applications, utility does not requires scale}

Different tasks can also call for models of different sizes, as seen on \autoref{fig:benchmarks}. Using these benchmarks as a reference for comparison, we see that e.g. a
1 GB model performs well on medical image segmentation, even though the images themselves are relatively large.
In computer vision, a 0.7 GB model can achieve good performance on a task such as object detection, even though scene parsing could require up to 3 GB of memory. For text processing, 
a natural language understanding task could be addressed with an LLM with hundreds of billions of parameters requiring more than 20 GB of memory (and multiple GPUs), even though a 1.3 GB model can also provide a good
semantic embedding, which can be leveraged for solving the same task.
Particularly for more focused tasks, smaller models often suffice: object
detection (\ref{fig:benchmarks}b) versus scene parsing (\ref{fig:benchmarks}c, or semantics (\ref{fig:benchmarks}e) versus 'understanding' (\ref{fig:benchmarks}f).

Given this varied landscape, and the particularities in size and performance across tasks, where should we direct our research efforts? There is no single answer to this question, but it is instructive that a proposal made by Wagstaff over a decade ago called for 
ML with meaningful applications, each of which calls for approaches adapted to its  constraints \cite{wagstaff2012machine}, a philosophy that has become decidedly unpopular in recent years in the pursuit of `general-purpose' AI models. Let us survey published applications of ML, to sheds some light on corresponding tradeoffs. 
We are slowly starting to see an opening towards ML applications driven by real-world needs of end users  -- for instance, in the \href{https://icml.cc/Conferences/2025/CallForPapers}{call for papers} for the main track of the 2025 ICML conference, which explicitly calls for ``Application-Driven Machine Learning" -- a direction typically relegated to less prestigious (and less selective) workshops in other ML conferences.

If we consider health applications, where we would like machine learning to  positively contribute to society, AI models are often built in data-scarce contexts
on which large models can more easily overfit \cite{varoquaux2022machine}. 
For example, to predict seizure outcome after epilepsy, \citet{eriksson2023predicting} find no
prediction-performance difference between logistic regression,
boosted trees, or deep learning, even as these approaches differ dramatically in terms of resources required. 
The largest source of medical data are probably electronic health records -- 
\citet{rajkomar2018scalable} reported performance of deep learning,
though their appendices show that logistic regression achieves the same performance. Later, \citet{yeche2021hirid} find deep-learning under-performing in a thorough electronic health records benchmark.
For education, one promise of machine learning is that it could produce customized pedagogy, shaped to the student based on which topics or skills they have learned, and which they are striving the understand. For this purpose, a random embedding of the student's learning sequence outperforms more resource intensive optimized deep or Bayesian models \cite{ding2019deep}.
For robotics, \citet{fu2024mobile} present an impressively useful system that can autonomously complete many daily-life tasks such as serving dishes whose key is co-training by a human. The learning systems used are rather modest, such as a ResNet18 architecture \cite{he2016deep}, dating from 2016 with 11 million parameters, which only take up 22 MB of memory with a precision of  bfloat16--which is dwarfed by many of today's SOTA systems, whose size is often measured in GB (\autoref{fig:benchmarks}).

If we inspect applications such as those mentioned above, we can see that for applied ML research what
often matters most -- even if we focus on simple purpose-specific metrics -- is to address well-defined and application-specific
goals while respecting the compute constraints at hand. And doing so often benefits from smaller, purpose-specific approaches as opposed to larger, more generic models. The benefit of purpose-specific models is also apparent in business applications: specificities of the business model lead to well-defined goals (efficiency, profit, etc.)~\citep{bernardi2019150}. Industry surveys indeed show that the most frequently used ML models are linear models and tree-based ones
\citep{survey2022}.

Happily for the AI research community, a focus on smaller models also unearths myriad unresolved scientific questions, which remain in need of scrutiny and innovative thinking independent of scale.
For example, for many applications, bottlenecks lie in decision-making, calibration, and
uncertainty quantification~\cite{van2019calibration}, not more compute or data.
Similarly, causal machine learning and distributional shift are very active areas of
research that are important irrespective of scale~\cite{kaddour2022causalmachinelearningsurvey}. And interpretability is still an unrealized goal for most neural-network-based approaches, even as the ability to understand and audit such systems remains centrally important in the context of meaningful real life applications, especially in high-stakes scenarios like health~\cite{murdoch2019definitions,ghassemi2021false}. 

\subsection{Unrepresentative benchmarks show outsize benefits from scale}
Progress in AI is measured by an ever-evolving set of benchmarks which are meant to enable comparing the performance of various models in a standardized way. In fact, one of the main reasons ML practitioners are drawn to scale is that large scale models have, over the past decade or more, beaten smaller models, as measured by these benchmarks. And beating the benchmark is currently how SOTA is defined (and how best paper awards are allotted, tenure given, funding secured, etc.). 
This persists even as this benchmark-centric evaluation often comes with ill-advised practices~\cite{flach2019performance}, both in terms of reproducibility of results~\cite{pineau2021improving,belz2021systematic,marie2021scientific} as well as the metrics used for measuring performance~\cite{post-2018-call,wu2016googles}. In fact, one problem that the ML community currently faces is that many benchmarks were created to assess the performance of models in the context of the academic research field, not as a stand in for measures of contextualized performance,  especially given the diversity of downstream contexts of applications in which AI models can be used, or to support broad claims made by marketers or companies given rising industrial interest and investment in AI. 

Furthermore, with the advent of generative models including LLMs, the ML community has faced new evaluation challenges given the openendedness of model outputs. It is no longer possible to simply evaluate prediction on labels in the test set when there is no canonical `correct answer'. This has resulted in a diversification of model evaluation practices, with the development of many different benchmarks intended to evaluative different aspects of generative model performance \citep{koch2021reduced,chang2023survey} -- this proliferation of non-standard approaches further confuses the landscape of assessment. To make matters worse, examining training datasets often reveals evidence of data contamination~\cite{dodge2021documenting}, including the presence of benchmark datasets inside training corpora~\cite{deng2023investigating}. This renders evaluations against these benchmarks questionable. All these shortcomings in benchmarks suggest that we should be very cautious regarding the validity of current evaluations of industrial models, given that the datasets used to train them are often not accessible to carry out the kind of public scrutiny necessary to identify such contamination or to understand the suitability of a given evaluation approach to the model at hand.

These issues have led to research on more rigorous model evaluation, including libraries \citep{eval-harness,von2022evaluate} and open leaderboards~\cite{open-llm-leaderboard}. However, the crux of the matter remains that ML benchmarks and evaluations remain unrepresentative proxies of downstream model performance, and are not a substitute for more holistic assessments. Importantly, such benchmarks cannot speak to whether a given model is suited to a given purpose and context. In addition, the “cost functions” used for many of these comparisons are not absolute and result in, e.g., the fixation on specific metrics such as accuracy or precision, while ignoring others, such as robustness or efficiency, which remain crucial in real-world applications~\cite{rogers2024position}. Finally, some ad-hoc benchmarks claim to measure ill-defined performance properties that are in fact impossible to verify (e.g. claiming that certain LLMs exhibit artificial general intelligence~\cite{bubeck2023sparks}). 
Considering a set of model capabilities much more narrow and better defined that general intelligence, model reasoning, thorough inspection of model capabilities has shown that benchmarks cannot be trusted as they are full of shortcuts exploited by models \cite{mirzadeh2024gsm}.
Such frail measures of high-level model capabilities impact the way the field as a whole tracks and perceives progress towards an ill-defined, illusory, and unattainable goal of general intelligence.

\section{Consequence 1: An unsustainable trajectory}
\label{sec:unsustainable}

As we have shown in the sections above, focusing all of our research efforts on the pursuit of bigger models narrows the field, and scale is not an efficient solution to every problem. There are many interesting problems in AI that do not require scale. The singular pursuit of scale also comes with worrying consequences, which we examine below.

\paragraph{Efficiency can have rebound effects}
As we have seen from historical data, the compute required to create and deploy SOTA AI models grows faster than the cost of compute decreases. And yet, if we succeed at making these technologies generally useful, deployment at massive scale would be the logical next step, making these SOTA models accessible to increasing quantities of users. We may hope that efficiency improvements will come to the rescue, but the recent history of AI suggests the opposite. 
It is well-known phenomenon in economics that when the efficiency of a general-use technology increases, the falling costs lead to an increased demand, resulting in an overall increase in resource usage. This is referred to as ``Jevons Paradox"~\cite{jevons1965coal}. Indeed, the AlexNet paper, which helped launch the current bigger-is-better era of AI research, was itself introducing a profound efficiency improvement, showing that a few GPUs could compete with tens of thousands of CPUs. One could assume that this would make AI training ``democratically accessible'', but in fact the opposite happened, with subsequent generations of AI models using even larger quantities of GPUs and training for a longer amount of time -- with the training of the most recent generations of LLMs being measured in \emph{millions} of GPU hours. 
This counter-intuitive dynamic has been documented numerous times in energy, agriculture, transportation, and computing~ \cite{york2016understanding}, where observed trends have led to truisms such as Wirth's law, which asserts that software demands outpace hardware improvements~\cite{wirth1995plea}.

\subsection{The costs of scale-first AI don't add up}

\begin{figure}[t!]
\centerline{\includegraphics[width=.7\linewidth]{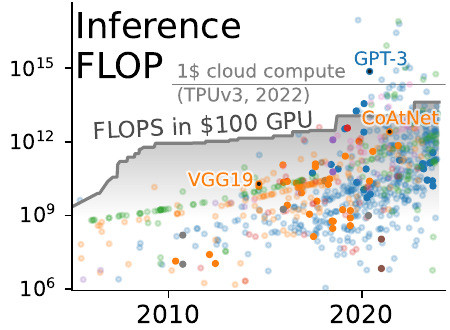}}
    \caption{\textbf{The cost of a single inference is growing faster than compute is improving}%
    \label{fig:inference-cost}}
\end{figure}

In fact, in many organizations, compute costs are already a major roadblock to developing and deploying ML models~\cite{paleyes2022challenges}. This is true even at large, tech-savvy companies such as Booking.com, who have found that gains in model performance cease to translate into gains in value to the company at a certain point, at which other factors such as efficiency or robustness become more important~\cite{bernardi2019150}.
Increase in model scale aggravates this problem: even after the costs of training are accounted for, large models come with large deployment costs. Indeed, the price of a single inference (i.e. model query) has been growing in the last two decades despite hardware efficiency improvements (\autoref{fig:inference-cost}).

Prohibitive costs of model deployment at scale threatens the business
model of AI. Due to these costs, major investors are questioning the
financial health of the current AI massive growth \cite{cahn2024ai,goldman2024ai}.
For instance, the compute required to run ChatGPT as a publicly accessible interface cost OpenAI \$\,700\,000 per day in 2023~\cite{etime2023bankruptcy}, and OpenAI's 2024 estimated balance is a \$\,5\,B loss
\cite{nytimes2024openai}. According to Alphabet's chairman, a search using Bard, which is powered by Google's recent PaLM 2 model~\cite{anil2023palm}, is 10 times more expensive than a pre-Bard Google search \cite{reuters2023bard}.
In a similar vein, \citet{bornstein2023owns} reports that in generative AI gross margins are ``more often as low as 50-60\%, driven largely by the cost of model inference''.
The problem of costs is increasingly pressing and many recent announcements around industry R\&D have focused on the cost or energy efficiency of new developments: OpenAI announced their recent GPT-4o mini as ``advancing cost-efficient intelligence'' \cite{openai2024gpt4o}. DeepSeek-v3 \cite{liu2024deepseek} is a very large (671B parameters) and capable language model which positions itself putting forward much decreased training and inference costs. The modernBERT paper \cite{warner2024smarter} tackles language processing while specifically avoiding generative modeling to decrease costs. Microsoft CEO Satya Nadella even argued in late 2024 that progress in generative AI should be measured in ``token per dollar per watt'', as opposed to metrics focusing only on accuracy- or cost~\cite{grallet2024nadella}.
As models are improving, the costs of one query are going down for a given model size. However, will this be enough? ``GenAI'' product development is calling not only for increased model scale --\emph{eg} to alleviate hallucinations--, but also increasing quantities of inferences by these models, adding them in user-facing applications on the web, but also in smartphones, smart connected devices, cars, etc.

Beyond generative AI, self-driving cars provide cautionary example. Here again the economics are challenging -- `autonomous' driving has turned out to rely heavily on the support of human workers and expensive sensors~\cite{wsj2022selfdriving}. As of 2024, the field of companies endeavoring to commercialize fully autonomous vehicles has shrunk considerably. Cruise suspended its efforts, leaving only Waymo remaining \cite{reuters2023cruise}.

\subsection{Worrying environmental impacts}

The emphasis on scale in AI also comes with consequences to the planet, since training and deploying AI models requires raw materials for manufacturing computing hardware, energy to power infrastructure, and water to cool ever-growing datacenters. The energy consumption and carbon footprint of AI models has grown with models such as LLMs emitting up to 550 tonnes of CO$_2$ during their training process~ \citep{strubell2019energy,luccioni2023counting,luccioni2022estimating} .
But the most serious sustainability concerns relate to inference, given the speed at which AI models are being deployed in user-facing applications. It is hard to gather meaningful data regarding both the energy cost and carbon emissions of AI inference because this varies widely depending on the deployment choices made (e.g. type of GPU used, batch size, precision, etc.). However, high-level estimates from Meta attribute approximately one-third of their AI-related carbon footprint to model inference \cite{wu2021sustainable}, whereas Google attributes 60\% of its AI-related energy use to inference~\cite{patterson2022}. Comparing estimates of the energy required for inference for LLMs~\cite{luccioni2024power} to ChatGPT's 10 million users per day~\cite{oremus2023ai} reveals that within a few weeks of commercial use, the energy use of inference overtakes that of training. Given the efforts to make AI more ubiquitous, these numbers could go up by many orders of magnitude. In fact, we are already seeing chatbots applied in consumer electronics such as smart speakers and TVs, as well as appliances such as refrigerators and ovens. The increased use of AI in user-facing products is putting tech companies' climate targets at risk, with both Microsoft and Google announcing in 2024 that they would miss the sustainability targets they set in previous years due to the energy demands of AI~\cite{rathi2024microsoft,metz2024google}.

\begin{figure}[t]
\centerline{%
    \includegraphics[width=.7\linewidth]{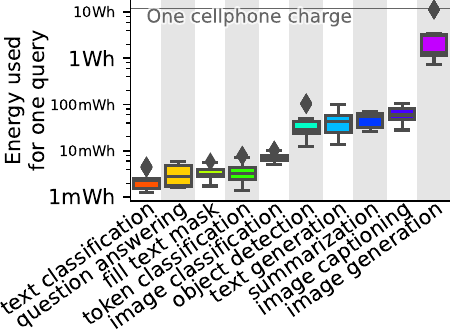}%
}
\caption{\textbf{A single inference uses more energy for models with broad purposes}.
 Data from \citet{luccioni2024power}.
    \label{fig:inference-energy}}
\end{figure}

Comparing different model architectures also shows that ``general-purpose'' (multi-task or zero-shot) models have higher energy costs than specialized models that were trained or fine-tuned for a specific task \citep[][and \autoref{fig:inference-energy}]{luccioni2024power}. This means that the current trend of using generative AI approaches for tasks such as Web search, which were previously carried out using ``legacy'' (information retrieval and embedding-based) systems, stands to increase the environmental costs of our day-to-day computer use significantly. 
The impact of a models' inference may seem a trifle compared to say a plane ride, however here again the challenge comes from the huge number of times they can be used. Data centers are already putting noticeable pressure on electricity networks of modern countries such as the US \cite{bloomberg2024ai}, leading tech companies to fund nuclear power plants for their exclusive usage \cite{npr2024threemile}.
Taking a step back, according to recent figures, global data centre electricity consumption represents 1-1.3\% of global electricity demand and contributes 1\% of energy-related greenhouse gas emissions~\cite{hintemann2022cloud,ict2020}. It is hard to estimate what portion of this number is attributable to AI. However, a recent report from the International Energy Agency estimates that electricity consumption from data centres and AI is set to double in 2 years, surpassing that of Japan (1 000 TWh) in 2026.~\cite{iea2024}.

\section{Consequence 2: More data, more problems} \label{sec:data}

\begin{figure}[b!]
    \centerline{\includegraphics[width=.7\linewidth]{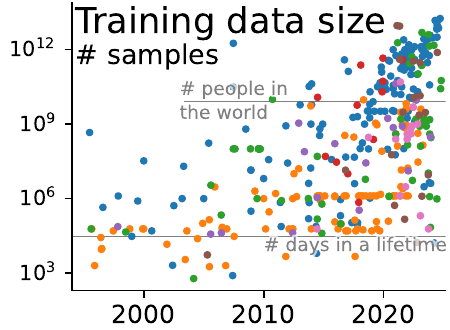}}
    \caption{\textbf{A sharp increase in amount of data used for
    training}
    Details in \autoref{app:historical_plots}.
    \label{fig:training_size}
    }%
\end{figure}

As ML datasets grow in size (\autoref{fig:training_size}), they bring a slew of issues ranging from documentation debts to a lack of auditability to biases. We discuss these below.

\paragraph{Data size in tension with quality}
In recent years, pretraining has become the dominant approach in both computer vision~\cite{szegedy2015going,redmon2016look} and natural language processing~\cite{devlin2018bert,liu2019roberta}. This approach requires access to large datasets, which have grown in size, from millions of images for datasets such as ImageNet~\cite{deng2009imagenet} to gigabytes of textual documents for C4~\cite{raffel2020exploring} and billions of image-text pairs in LAION-5B~\cite{schuhmann2022laion}. The premise of pretraining is that more data will improve model performance and ensure maximal coverage in model predictions, with the assumption that the more data used for pretraining, the more representative the model's coverage will be of the world at large. However, numerous studies have shown that neither image nor text datasets are broadly representative, reflecting instead a select set of communities, regions and populations~\cite{dodge2021documenting,rogers2021changing,raji2021ai,luccioni2023bugs}. In fact, recent research on the LAION datasets showed that as these datasets grow in size, the issues that plague them also multiply, with larger datasets contain disproportionally more problematic content than smaller datasets~\cite{birhane2023into}. \citet{thiel2023identifying} reported that the LAION datasets included child sexual abuse material, prompting the datasets to be taken down from several hosting platforms. But with dozens of image generation models trained on LAION variants, some already deployed in user-facing settings, it is hard to assess and mitigate the negative effects of this grim reality~\cite{sambasivan2021everyone}. 

While a growing wave of research has proposed a more `data-centric' approach to data collection and curation~\cite{zha2023data,mitchell2022measuring}, attempts to actually document the contents of ML datasets have been hampered by their sheer size, which requires compute and storage beyond the grasp of most members of the ML research community~\cite{luccioni2021s}, before the challenges of classifying and understanding the contents of a given dataset are even addressed. As the field works with ever larger datasets, we are also incurring more and more `documentation debt' wherein training datasets are too large to document both during creation and post-hoc~\cite{bender2021dangers}. In a nutshell, this means that we do not truly know what goes in to the models that we rely on to answer our questions and generate our images, apart from some high-level statistics. And this, in turn, hampers efforts to audit, evaluate, and understand these models.

\paragraph{Invasive data gathering}
From a privacy perspective, the perceived need for ever-growing datasets implicitly incentivizes more pervasive surveillance, as companies gather our clicks, likes and searches to feed models that are then applied to sell us consumer products or impact the order the search results that we see. In fact, the majority of the revenue of the Big Tech companies leading the AI revolution comes from targeted advertising - representing 80\% of Google's \cite{form2023google} and 90\% of Meta's \cite{form2023meta,report2023meta} annual revenue in 2022. 

Also, while much of ML data gathering efforts have been operating under the assumption that copyright laws do not apply for data gathered from the Internet and used to train ML models (or, at the least, that training ML models constitutes `fair use'), last year has seen a series of copyright lawsuits filed against many companies and organizations. This includes lawsuits from authors, artists and newspapers \cite{small2023silverman,schrader2023class,grynbaum2023times}. While is is too early to tell what the verdicts of these lawsuits will be,  they will undoubtedly impact on the way in which ML datasets are created and leveraged. The European Union's General Data Protection Regulation (GDPR) law sets some limitations in terms of data gathering and privacy in the EU, even if unevenly enforced. However, the United States and many other jurisdiction have yet to pass federal privacy laws, offering their citizens little to no privacy protection against predatory data gathering practices and surveillance, which the rush to scale AI incentivises.

\section{Consequence 3: Narrow field, few players}
\label{sec:narrowing}

\subsection{Scale shouldn't be a requirement for science}

The bigger-is-better paradigm shapes the AI research field, yet we are in a moment where expensive and scarce infrastructure is viewed as necessary to conduct cutting-edge AI research and where companies are increasingly focused on providing resources to large-scale actors, at the expense of academics and hobbyists~\cite{mossnvidia}. Recent research argues that “a compute divide has coincided with a reduced representation of academic-only research teams in compute intensive research topics, especially foundation models.”~\citep[p.1]{besiroglu2024compute} We also see evidence of this in the way in which the AI field is both growing--as measured by number of PhDs in AI, and shrinking--as measured by graduates remaining in academia. \citet{stanford2023ai} reports that “the proportion of new computer science PhD graduates from U.S. universities who specialized in AI jumped to 19.1\% in 2021, from 14.9\% in 2020 and 10.2\% in 2010.”. This dynamic sees academia increasingly marginalized in the AI space and reliant on corporate resources to participate in large-scale research and development of the kind that is likely to be published and recognized. And it arguably disincentives work that could challenge the bigger-is-better paradigm and the actors benefiting from it.

\subsection{Scale comes with a Concentration of power}

The expense and scarcity of the ingredients needed to build and operate increasingly large models benefits the actors in AI that have access to compute and distribution markets. This works to concentrate power and influence over AI, which, in turn, provides incentives for those with sufficient resources to perpetuate the bigger-is-better AI paradigm in service of maintaining their market advantage. 

\paragraph{The scale game benefits large players}
Market concentration is apparent in compute resources, and is most evident when examining Nvidia GPUs. The cost and scarcity of these resources make it increasingly difficult for an academic lab, or even most startups, to purchase and rack sufficient hardware on premises. An H100, currently Nvidia’s most powerful chip, costs up to \href{https://www.tomshardware.com/tech-industry/artificial-intelligence/nvidias-h100-ai-gpus-cost-up-to-four-times-more-than-amds-competing-mi300x-amds-chips-cost-dollar10-to-dollar15k-apiece-nvidias-h100-has-peaked-beyond-dollar40000}{\$40,000}. Recent shortages in chips have intensified the gap between the ``GPU rich'' and ``GPU poor''~\cite{barr2023gpu} -- notably, this constrained supply has benefited the large cloud providers who have privileged access to such hardware~\cite{bornstein2023owns}. 

\begin{figure}[t!]
    \centerline{\includegraphics[width=.7\linewidth]{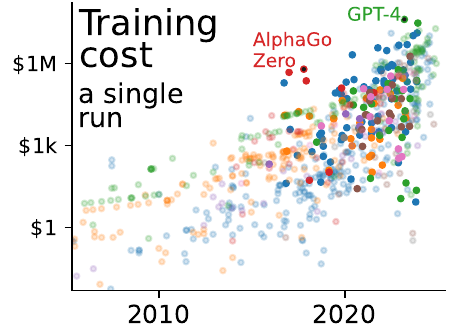}}
    
    \caption{\textbf{The increase in training costs leaves many aside}
    \label{fig:compute_cost}}
\end{figure}

For training and deploying large models, most AI practitioners operating outside of large cloud companies must rent access to compute from these companies. Since the cost of training ML systems has grown exponentially, this can entail costs for large scale models in the tens to hundreds of millions of dollars, which is beyond the budget of the vast majority of academic researchers (\autoref{fig:compute_cost}). However, tech companies are more able to invest this money to fuel their AI research and competitiveness -- Meta alone is estimated to have spent \$18 billion on GPUs in 2024~\cite{hays2024}, while OpenAI announced early 2025 StarGate, a joint venture with SoftBank, Oracle, and MGX, planning more than \$ 100 billion AI-infrastructure investments \cite{openai2024stargate}, bigger than the GDP of most countries in the world.
While AI startups continue to raise significant capital, the majority of this capital (“up to 80-90\% in early rounds”) is paid to cloud providers~\cite{bornstein2023owns}. Notably, the flow of capital here is often circular: three large cloud companies ``contributed a full two-thirds of the \$27bn raised by fledgling AI companies in 2023'' \cite{ft2023outspends}. 

These circular investment deals can blur the boundary between “investment” and “acquisition”. For instance, the deal between Microsoft and OpenAI \citep[currently under investigation by the FTC,][]{ftc2024}, involved providing OpenAI access to Azure compute, in exchange for exclusive license to integrate OpenAI’s GPT models, and a promise of \$1 trillion in profit delivered to Microsoft prior to any revenue being directed to OpenAI’s social mission.

\paragraph{Shaping practices and applications}
AI dominance is one factor that has helped vault companies into shapers of global economic forecasts and markets. The S\&P 500, used by US investors as an index of market performance, is significantly shaped by the fates of large US tech companies~\cite{rennison2023big}. This indicates that the interests of large AI companies, and particularly their emphasis on large-scale AI, have implications well beyond the tech industry, since a downturn in the AI market would impact global financial markets beyond. This creates market incentives, divorced from scientific imperatives, for perpetuating the bigger-is-better paradigm on which these companies are betting. 

The expense of creating large-scale AI models also makes the need to commercialize these models more pressing -- even if inference can be costly~\cite{vanian2023chatgpt}. Here, established cloud and platform companies also have an advantage in the form of access to large and existing markets. Cloud infrastructure offerings are a readily available means to commercialize AI models, allowing startups and enterprises to rent access to, e.g., generative AI APIs as part of cloud contracts. The ecosystem of `GPT wrapper' companies licensing access to the model via Microsoft's Azure cloud is an obvious example. \citet{mckinsey2022state}, looking at the 2022 landscape of AI, noted that ``commercialization is likely tied to hosting. Demand for proprietary APIs (e.g. from OpenAI) is growing rapidly.'' 
Notable here is that the business model for commercializing generative AI in particular is still being worked out \cite{bornstein2023owns}.

\paragraph{Consequences for innovation}

When a few actors control an economy, it stifles innovation, as illustrated by recent history of computing.
The 1990s were marked by the explosion of personal computing, and the race to commercialize networked computation; this innovation reshaped the economy and the labor market \cite{caballero2010creative}. From there, the actors involved became concentrated into a small number of firms, becoming enablers and gatekeepers of personal-computing software. Resulting market capture created a rent, to the expense of the rest of the economy, including IT consumers
\cite{aghion2023}. This, in turn, led to a fall of innovation and growth because of lack of incentives for leaders and lack of market access for competitors \cite{aghion2023}; desktop software has been moving slowly.

Today we face a similar scenario. Even the most well-resourced AI startups--like OpenAI, Anthropic, or Mistral-- currently need compute resources so large that they can only be leased from a handful of dominant companies. And the primary pathway to a more widespread commercialization of AI models also lies through these companies.
For example, Microsoft licensed their GPT API to `wrapper' startup Jasper before they launched ChatGPT -- Jasper provided writing help, offering services very similar to ChatGPT \cite{pardes2022unicorn}. This, and OpenAI's launch of their GPT app store, inflamed concerns that Microsoft and OpenAI would leverage their informational advantage and privileged ability to shape the AI market and to compete with smaller players dependent on their platform. This echos practices that Amazon Marketplace has been criticised for, in which they used data collected about buyers and sellers to inform product development that directly competed with marketplace vendors \cite{wsj2020amazon}.

\paragraph{Scale comes with consequences to society}

The growth and profit imperative of corporations dictates finding sources of revenue that can cover the costs of AI systems of ever-growing size. These often puts them at odds with responsible and ethical AI use, which puts the emphasis on more consensual and incremental progress. We see this dynamic at play in OpenAI’s shifting stance and policies as they have become more closely tied to Microsoft, a choice they made admittedly to access scarce computational resources. This is illustrated by the 2024 change to their acceptable use policy, which significantly softened the proscription against using their products for ``military and warfare''; this change unlocked new revenue streams \cite{times2024military}, connected OpenAI's large scale AI to Microsoft's existing US military partnerships \cite{vice2023defense}, and later pushed AI to direct battlefield application via a partnership with the defense-tech company Anduril \cite{odonnell2024openai}. 
The public did not have a say in this determination, nor, we assume, did the global stakeholders whose interests may not be safeguarded by the militaries that OpenAI chooses to provide services to. And yet, the concentrated private industry power over AI creates a small, and financially incentivized segment of AI decision makers.  We should consider how such concentrated power with agency over centralized AI could shape society under more authoritarian conditions. 

There are many other examples of financial incentives shaping the use of automated decision systems in ways that are not socially beneficial. \citet{ross2023unitedhealth} showed United Health used a diagnostic algorithm to deny patients care, an application now at the center of a class action lawsuit filed by patients and their families \cite{us2023classaction}. The bigger-is-better AI paradigm is also exacerbating geopolitical concentration and tensions. Beyond the power of large corporations, government imperatives are also in play, particularly given that the large AI industry players are primarily located in the US. These tensions are apparent in the efforts on the part of the US government to limit the supply of computing resources such as GPUs to China in the global race for AI dominance \cite{reuters2023china}. While a discussion of whether such limits are warranted is outside the scope of this paper, this dynamic highlights the role of AI as a source of geopolitical power, one that given the current concentration of AI resources threatens to undermine technological sovereignty globally. 

\paragraph{A broader landscape of risks} Deploying AIs is society may come with a variety of risks \cite{bengio2024international}. Some risks are already visible in existing systems such as the problems of fairness and replicating biases which lead to underdiagnosis in historically under-served populations \cite{seyyed2021underdiagnosis}. Other risks are more speculative given today's systems but are to be considered in future scenario where AIs become much more powerful \cite{critch2023tasra}. A concentration of power fostered by very large-scale AI systems that can be operated and control by only very few actors makes these multiple risks more problematic, as it challenges checks and balance mechanisms. 

\section{Ways Forward: small can also be beautiful}

Machine learning research is central to defining technical possibilities and building the narrative in AI. There is no magic bullet that unlocks research on AI systems both small-scale and powerful, but shared goals and preferences do shape where research efforts go. We believe that the research community can and must act to pursue scientific questions beyond ever-larger models and datasets, and needs to foster scientific discussion engaging the trade-offs that come with scale. As such, we call the research community to adopt the following norms:

\paragraph{Assigning value to research on smaller systems} By defining research agendas (which papers and talks are accepted, which questions are asked), the research community can shape how AI progresses and is appreciated. We can deepen our understanding of performance by diversifying our benchmarks, investigating the limitations of these benchmarks, and building bridges to other communities to pursue tasks and problems that are relevant in different contexts. All work on large AI system should also be compared to simpler baselines, to open the conversation on trade-offs between scale and other factors. And we should value research on open research questions such as uncertainty quantification or causality, even when it is conducted on smaller models, as it can bring valuable insights for the field as a whole. 

\paragraph{Talking openly about size and cost} A scientific study in machine learning should report compute cost, energy usage and memory footprint for training and inference. Measuring these requires additional work, but it is small compared to the work of designing, training, and evaluating models, and can be done with available open-source tools like \href{https://www.tensorflow.org/tensorboard}{TensorBoard} and \href{https://codecarbon.io/}{Code Carbon}. We should account for efficiency as well as performance when comparing models for instance, reporting metrics such as samples/kWh or throughput.

\paragraph{Holding reasonable expectations} Experiments at scale are costly, and we cannot expect everyone in the ML community to be ``GPU rich''. When assessing the merits of a scientific study, we should refrain from requiring additional experiments more costly than those already performed. And, as ML researchers, we should keep in mind that \emph{1)} not every problem is solved by scale \emph{2)} solving a problem that is only present at small scales is valuable as it decreases costs.

\bigskip

\begin{figure}[t!]
    \centerline{\includegraphics[width=.82\linewidth]{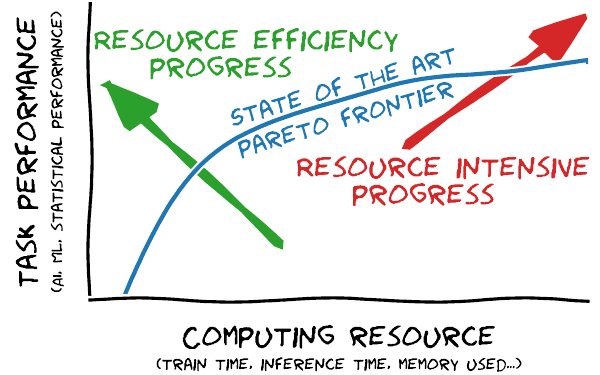}}
    \caption{{\bf Pareto optimality and different contributions} The state-of-the-art should be considered as a Pareto frontier in the trade-off space between task performance and computing resource. Visible contributions are often ``resource intensive progress'', increasing task performance by increase computing resources. We must also celebrate resource efficiency progress that decrease computing resources for the same task performance.
    \label{fig:pareto}}
\end{figure}

Quantitatively framing the changes that we call for can help anchoring them in the empirical practices of the AI community. For this, it is useful to consider, as in \autoref{fig:pareto}, the trade-off between task performance (e.g. performance on an ML benchmark) and the computing resources used. In this space, the state of the art appears as a pareto-optimal frontier. Successes that typically make the headlines are ``resource intensive'': they move the field in the upper right direction, where using more computing resources is associated to better task performance. We must also celebrate progress in resource efficiency, moving the Pareto frontier to the right: less resource for the same performance. We believe that this framing and the discussions it entails are useful to build the discussion, and we advocate their more systematic use by AI practitioners.

\paragraph{Conclusion}

As members of the AI community, we find that, in recent years, AI research has acquired an unhealthy taste for scale. This comes with dire consequences-- economic inequalities and environmental (un)sustainability, datasets that erode privacy and emphasize corrosive social elements, a narrowing of the field, and a structural exclusion of small actors such as most academic labs and many startups. This fixation on scale has emerged via norms that shape how the scientific community acts. We believe that scientific understanding and meaningful social benefits of AI will come from de-emphasizing scale as a blanket solution for all problems, instead focusing on models that can be run on widely-available hardware, at moderate costs. This will enable more actors to shape how AI systems are created and used, providing more immediate value in applications ranging from health to business, as well as enabling a more democratic practice of AI.

\bibliography{biblio}
\bibliographystyle{ACM-Reference-Format}

\newpage
\appendix
\onecolumn
\section{Methodology for the plots on historical trends}%
\label{app:historical_plots}

Here we give details on the plots on historical trends 
(\autoref{fig:trends_scale}, \autoref{fig:inference-cost},
\autoref{fig:compute_cost}, \autoref{fig:training_size}). All the Python
code to generate the plots is available as supplementary material and
will be deposited on github.

\subsection{Sources of data}

\paragraph{Evolution of notable AI systems} The data on the notable AI systems
is adapted from \citet{epochMachineLearningData2022}
(\url{https://epochai.org/data/epochdb/table}). Here, a system is considered
notable if it is highly cited. The data has been collected manually from
publications.

\paragraph{Cost of memory}
We used data on the cost of memory from OurWordInData:
\url{https://ourworldindata.org/grapher/historical-cost-of-computer-memory-and-storage}

\paragraph{Cost of GPU FLOPs} The cost of GPU FLOPs (FLOPs denote the number of
floating point operations per second) can be found here:
\url{https://epochai.org/blog/trends-in-machine-learning-hardware}

An estimate of the cost of compute with TPUv3 can be found here:
\url{https://blog.heim.xyz/palm-training-cost}: in 2022 ``\emph{We can rent a
TPUv3 pod with 32 cores for \$32 per hour. So, that's one TPUcore-hour per
dollar.}''

\paragraph{Largest supercomputer} Data on the computing power of the largest
super computer on Earth can be found here:
\url{https://en.wikipedia.org/wiki/File:Supercomputers-history.svg}

\paragraph{FLOPs of other computer} Data on the computing power of 
other computers, including the PlayStation 4 can be found here:
\url{https://en.wikipedia.org/wiki/FLOPS}

\subsection{Imputation of missing values}

Not all features of notable AI systems are reported, for instance
inference cost, compute cost in dollars, may often be missing. For all
plots but the one about dataset size\footnote{We did not impute the
dataset size as, unlike the other quantity, it does not need to have
links to other quantities.} we impute the missing quantities from the
reported one. Specifically, we use an imputation with chained equations
\citep[MICE]{van2011mice}. We relate number of parameters, training FLOP,
inference FLOP, training cost, to each other as well as to dataset size,
number of epochs, application domain, and time the model was published.

The imputed data points are displayed on the various figures with
transparency.

\subsection{Estimation of doubling time}%
\label{sec:doubling_time}

\begin{figure}[]
    \begin{minipage}{.41\linewidth}
    \caption{\textbf{Doubling time of models}, estimated between years
    2018 and 2024, for the average model, and the $95^{th}$ percentile.
    Data from \citet{epochMachineLearningData2022}; the estimation is detailed in \autoref{sec:doubling_time}.%
    \label{fig:doubling_time}%
    }
    \end{minipage}%
    \hfill
    \begin{minipage}{.55\linewidth}
	\includegraphics[width=\linewidth]{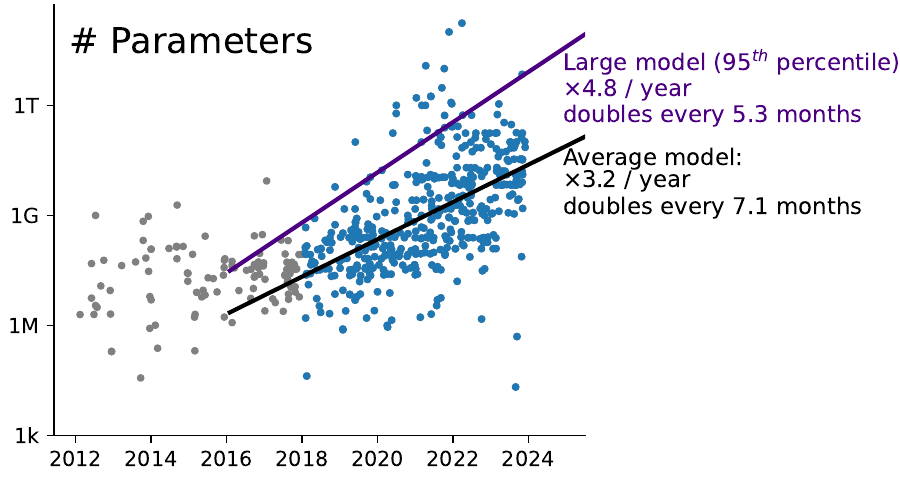}%
    \end{minipage}
\end{figure}

We estimate the doubling time of the number of parameters for the span
2018..2024. For this we use a linear model in log space: an ordinary
least square regression to compute the evolution of the average model
size as a function of time, and a quantile regression (pinball loss) to
compute the evolution of the large models ($95^{th}$ percentile of the
size distribution). We find a doubling time of 7.1 months for the average
model, which corresponds to a yearly increase of $\times 3.2$ per year,
and a doubling time of 5.3 months for the large models, which corresponds
to an increase of $\times 4.8$ per year (\autoref{fig:doubling_time}).

The corresponding increase is very rapid: in three years, the size large
models is multiplied by 110, and that of average models by 33. If growth
continues at this pace, in 10 years large models will be 6.5 million
times bigger, and average models 100\,000 times bigger.

\section{Methodology for the plots on benchmarks}%
\label{app:benchmark_plots}

Here we give details on the plots on the benchmarks
(\autoref{fig:benchmarks}).

\subsection{Sources of data}

The difficulty with doing an analysis of how performance relates to scale
is that they few benchmarks report a measure of scale. We used the
following sources:

\paragraph{Tabular data} The data on tabular learning comes from the
benchmark of \citet{grinsztajn2022tree}, with detailed results downloaded
from
\url{https://figshare.com/ndownloader/files/40081681}. The timings and
performance results were measured by \citet{grinsztajn2022tree} by
running the algorithms for their benchmark.

\paragraph{Medical imaging} The data on medical imaging comes from an
organ segmentation challenge: the MICCAI FLARE23 challenge (Fast,
Low-resource, and Accurate oRgan and Pan-cancer sEgmentation in Abdomen
CT):
\url{https://codalab.lisn.upsaclay.fr/competitions/12239#learn_the_details-testing_results}
The memory result was OCRed from the image of the table on the page. As
the table report area under the GPU memory curve (as measured by the
challenge organizers), we divided by time to have average GPU memory.

\paragraph{Object detection, COCO data} We consider the coco
computer-vision object detection benchmark \citep{lin2014microsoft}
and retrieve benchmark results from
\url{https://paperswithcode.com/sota/object-detection-on-coco}. To have a
model size in RAM (to compare with other benchmarks) we assume that
parameters are bfloat16, which is on the low end of the memory footprint.

\paragraph{Scene parsing, ADE20K} We consider the ADE20K computer-vision
scene parsing \citep{zhou2017scene} and retrieve benchmark results from
\url{https://paperswithcode.com/sota/semantic-segmentation-on-ade20k}.
Here also we extrapolate memory footprint from number of parameters
assume bfloat16 representation.

\paragraph{Massive Text Embedding Benchmark} We consider the massive text
embedding benchmark \citep{muennighoff2022mteb}, downloading results from 
\url{https://huggingface.co/spaces/mteb/leaderboard}.

\paragraph{LLM benchmark} We consider the ``open LLM leaderboard''
 \citep{open-llm-leaderboard}, downloading results from
\url{https://huggingface.co/spaces/HuggingFaceH4/open_llm_leaderboard}.

\subsection{Fit of the $.75$ quantile}

To study the dependence between model size and performance, we fit the
$.75$ quantile: the goal is to look at the ``best'' model for a
given parameter size, however given the scarcity of data we use the 
$.75$ quantile as conditional $\alpha$-quantiles are hard to estimate for
high $\alpha$ values.

We use a order-3 polynomial expansion to capture non-linear trends, and a
quantile regression. We estimate prediction intervals by bootstrap.

\section{Labor and the human cost}\label{app:labor}

While we, as a computer-science community, are well attuned to computational costs, other costs come in when developing large AI models. In particular,
recent breakthroughs in ML rely on significant human labor to create data at various points throughout the development and deployment life cycle. 
ImageNet, the dataset that helped catalyze the current AI era, was only possible due to significant click worker labor, namely Amazon Mechanical Turk, which replaced costly and slow student labor~\cite{feifei2018,deng2009imagenet}. Subsequently, the Mechanical Turk platform, and the ability it gave companies and researchers to remotely direct large numbers of low paid workers, became critical to construct emblematic computer vision datasets like MS-COCO~\cite{lin2014microsoft} and Flickr-30K~\cite{plummer2015flickr30k}. More recently, reinforcement learning from human feedback (RLHF) has become a central to conversational large language models, allowing them to better respond to human preferences~\cite{ziegler2020finetuning} and to provide outputs that stay within the boundaries of politically and socially acceptable expression. In fact, RLHF was hailed as the 'secret ingredient' of ChatGPT~\cite{metz2023secret}. RLHF continues the legacy of Mechanical Turk, paying workers, often in developing countries, a few dollars an hour for labor that is repetitive, exhausting, and often emotionally taxing~\cite{hao2023cleaning}. Even though each worker is paid little, given the scale of RLHF endeavors, it remains inaccessible to average research labs and startups, who cannot afford to pay the thousands of people required to carry out this 
tuning~\cite{widder2023open}.


\end{document}